\documentclass[12pt]{article}
\usepackage{graphicx,amssymb}
\def\be{\begin{equation}}
\def\ee{\end{equation}}
\def\bea{\begin{eqnarray}}
\def\eea{\end{eqnarray}}

\def \b  {\beta}
\def \pa  {\partial}
\def \f  {\frac}
\def \ell {\lambda}

\begin{document}

\begin{titlepage}

\begin{center}

\vskip 1.5 cm {\Large \bf Rotating Rindler-AdS Space \\[4 mm]}
\vskip 1 cm {Maulik Parikh\footnote{maulik.parikh@asu.edu}$^{a}$, Prasant Samantray\footnote{prasant.samantray@asu.edu}$^{b}$,
and Erik Verlinde\footnote{ e.p.verlinde@uva.nl}$^{c}$}\\
{\vskip 0.75cm $^{a}$ Department of Physics and Beyond: Center for Fundamental Concepts in Science\\
Arizona State University, Tempe, Arizona 85287, USA}
{\vskip 0.75cm $^{b}$ Department of Physics\\ 
Arizona State University, Tempe, Arizona 85287, USA} 
{\vskip 0.75cm $^{c}$ Institute for Theoretical Physics and GRAPPA\\
University of Amsterdam, Science Park 904, 1090 GL, Amsterdam, The Netherlands}
\end{center}

\vskip  .25 cm

\begin{abstract}
\baselineskip=16pt
\noindent
If the Hamiltonian of a quantum field theory is taken to be a timelike isometry, the vacuum state remains empty for all time. We search for such stationary vacua in anti-de Sitter space. By considering conjugacy classes of the Lorentz group, we find interesting one-parameter families of stationary vacua in three-dimensional anti-de Sitter space. In particular, there exists a family of rotating Rindler vacua, labeled by the rotation parameter $\beta$, which are related to the usual Rindler vacuum by non-trivial Bogolubov transformations. Rotating Rindler-AdS space possesses not only an observer-dependent event horizon but even an observer-dependent ergosphere. We also find rotating vacua in global AdS provided a certain region of spacetime is excluded.
\end{abstract}

\end{titlepage}

\section{Introduction}

In a theory with diffeomorphism symmetry, the choice of time, and therefore the choice of the time-evolution operator, is essentially arbitrary \cite{Davies}. However, if the Hamiltonian is chosen to be a timelike isometry, then the resulting time-evolution is special: energy is conserved and the vacuum remains empty. 

Here we consider the isometries of anti-de Sitter space in order to find stationary vacuum states. We find that there exists a one-parameter family of rotating stationary vacuum states in three-dimensional anti-de Sitter space. These exotic vacuum states, which we call $\beta$-vacua after the rotation parameter $\beta$, are inequivalent to the familiar vacuum states of these spacetimes, in that they are related to them by non-trivial Bogolubov transformations.

Intuitively, the reason for the existence of rotating vacuum states in anti-de Sitter space is the following. The isometry group of AdS can be regarded as the Lorentz group of a higher-dimensional flat embedding space, albeit with two time coordinates. The Lorentz group however has a rich structure: besides rotations and boosts, it also contains linear combinations of rotations and boosts that cannot be reduced by Lorentz transformations to either pure rotations or pure boosts. The main idea of this paper is to let the Hamiltonian be one of these timelike ``rota-boosts"  in order to obtain novel stationary rotating vacua.

\section{Stationary Vacua}

To find the possible stationary vacua of a given spacetime, we require three conditions to hold.

$\bullet$ The candidate Hamiltonian should of course be a continuous isometry.

$\bullet$ There must exist a region of spacetime that admits a Cauchy surface such that the Hamiltonian is future-directed and timelike at the (possibly asymptotic) spatial boundary of the region.

$\bullet$ The orbits of the Hamiltonian must not exit that region.

The motivation for these conditions is the following. The isometry may not be globally timelike, so we may have to restrict our quantum field theory to some region of spacetime, such as the static patch of de Sitter space or the Rindler wedge in Minkowski space. That region of spacetime should be globally hyperbolic (i.e. admit a Cauchy surface) so that time evolution of quantum states can be defined. For the same reason, the orbits of the candidate Hamiltonian should not exit the region. Finally, for the isometry to even be considered a Hamiltonian, it better be timelike at least at the asymptotic boundary of the region; we do not require the stronger condition that the Hamiltonian be timelike everywhere within our region so as not to preclude the existence of an ergosphere.

There may be more than one Hamiltonian that satisfies the above conditions. If the different possible Hamiltonians are isometrically equivalent i.e. if they can be related by isometries (so that they are both elements of the same conjugacy class of the isometry group), then they lead to the same vacuum state. However, if the Hamiltonians are isometrically inequivalent (being part of different conjugacy classes), then, given some quantum field theory, they could correspond to different vacuum states. 

To illustrate these ideas, let us find all the stationary vacua of Minkowski space \cite{Pfautsch}. The most general continuous isometry of Minkowski space is generated by a linear combination of translations, boosts, and rotations:
\be
\alpha^\mu P_\mu + \beta^i K_i + \omega^{ij} J_{ij} \; .
\ee
This must be timelike, at least in some suitable region, for the generator to be a candidate Hamiltonian. Choosing the Hamiltonian to be $P_0$ yields the usual Poincar\'e-invariant vacuum.  Next, we note that the boost generator, $K_i$, squares to $X_0^2 - X_i^2$, which is timelike when restricted to the wedges $X_i^2 > X_0^2$ and is future-directed when further restricted to $X_i > 0$. This is of course the right Rindler wedge. Moreover, the orbit of $K_i$ starting from a point in the right Rindler wedge remains in the wedge. Hence, $K_i$ is a candidate Hamiltonian for a stationary vacuum; indeed, choosing the Hamiltonian to be $K_i$ yields the Rindler vacuum for the right Rindler wedge, while choosing the Hamiltonian to be $-K_i$ gives the Rindler vacuum for the left Rindler wedge. It is straightforward to check that there are no other inequivalent isometric Hamiltonians. For example, the combination $P_0 + \omega J_{12}$, which generates the worldlines of observers rotating in the $X_1-X_2$ plane with angular velocity $\omega$, becomes spacelike outside the sphere $X_1^2 + X_2^2 = 1/\omega^2$ \cite{daviesrotating}; restriction to the world-volume of the inside of the sphere fails because such a region does not admit a Cauchy surface. Another possibility, the combination $P_0 + \beta K_i$, generates the worldlines of Rindler observers in a translated Rindler wedge. Or, the combination $P_0 + \alpha^i P_i$ is timelike for $\alpha^i \alpha_i < 1$ but this is obviously isometrically-equivalent to the Poincar\'e Hamiltonian via a Lorentz boost. It is easy to check that there are no other inequivalent isometries that could be used as the Hamiltonian.

Thus, in Minkowski space, the only stationary vacua corresponding to isometric Hamiltonians are the Poincar\'e-invariant vacuum and the Rindler vacuum.

\section{Conjugacy Classes of the Lorentz Group}

As we shall show, anti-de Sitter space permits a richer set of possibilities for stationary vacua. AdS space can be viewed as a hyperboloid embedded in a flat embedding space; the isometry group is therefore a higher-dimensional Lorentz group.

The Lorentz group has an interesting structure. There are five types of Lorentz transformations; that is, group elements of $SO(1,3)$ fall into five conjugacy classes. One conjugacy class consists of the elliptic transformations. This is the set of Lorentz transformations conjugate to the pure rotations i.e. the elliptic transformations consist of all Lorentz transformations, $\Lambda J_i \Lambda^{-1}$, that can be obtained from pure rotations via Lorentz transformations. Another  conjugacy class is that of the hyperbolic transformations; these consist of the pure boosts and their conjugates, $\Lambda K_i \Lambda^{-1}$. There is also the class of parabolic transformations, whose representative elements are the so-called null rotations, generated by $J_i + K_j$ for $i \neq j$. Most interesting for our purposes is the class of loxodromic transformations.\footnote{The peculiar names of the conjugacy classes are derived from  types of curves on a sphere, as named by maritime navigators. Because Lorentz transformations leave light cones invariant, the celestial sphere of an observer's night sky is mapped to itself. The orbits of the Lorentz transformations are curves on the sphere; a loxodrome (also known as a rhumb line) is a curve that spirals from one pole to the other while intersecting all longitudinal meridians at the same angle.} These are Lorentz transformations generated by a commuting pair of a rotation and a boost, such as $K_z + \beta J_z$. Rota-boosts cannot be reduced to either pure rotations or pure boosts by Lorentz transformations because obviously those lie in different conjugacy classes. Indeed, strictly speaking, the number of conjugacy classes is infinite, with each class labeled by a different value of the rotation parameter $\beta$. These are all the nontrivial conjugacy classes of $SO(1,3)$. (There is also the trivial conjugacy class containing the identity transformation.)

There is an electromagnetic analogy to the rota-boosts. The Lorentz generators $M_{\mu \nu}$, which are anti-symmetric, can be thought of as the electromagnetic field strength, $F_{\mu \nu}$; the boosts are then like the electric field and the rotations like the magnetic field. Then we know that there are Lorentz invariants of the type $F \wedge {}^* F \sim E^2 - B^2$ but also of the type $F \wedge F \sim E \cdot B$. If $E \cdot B \neq 0$, no Lorentz transformation can transform the field into a configuration that is either a pure electric field ($E^2 - B^2 > 0$), a pure magnetic field ($E^2 - B^2 < 0$), or pure electromagnetic ``radiation" ($E^2 - B^2 = 0$), since these all have $E \cdot B = 0$. Correspondingly, rota-boosts are generated by generators that have $J \cdot K \neq 0$.

Specifically, a generator of a rota-boost takes the form
\be
M_{01} - \beta M_{23}	\label{rotaboost}
\ee
in Cartesian coordinates. The key property is that rota-boosts are linear combinations of the usual Lorentz generators with no shared indices. In higher dimensions, there are additional parameters. For example, in six spacetime dimensions, there are two-parameter generators of the form
\be
M_{01} - \beta_1 M_{23} - \beta_2 M_{45}	\label{6Dlox}
\ee
The Lorentz-invariant Casimir which generalizes $J \cdot K$ is
\be
\epsilon_{i_1 \ldots i_{d}} \omega^{i_1 i_2} \ldots \omega^{i_{d-1} i_d} \; , \label{Casimir}
\ee
where $\omega$ is the parameter for the most general generator $\frac{1}{2} \omega_{ij}M^{ij}$. For example, the invariant of the generator (\ref{rotaboost}) is $2 \beta$.
In odd dimensions no invariant can be formed using the Levi-Civita tensor but it is nevertheless possible to argue that linear combinations of Lorentz generators with no shared indices cannot be reduced to elliptic, hyperbolic, or parabolic transformations. We will see that taking the Hamiltonian to be a generator of rota-boosts leads to novel stationary vacua in three-dimensional anti-de Sitter space.

\section{Anti-de Sitter $\beta$-Vacua}
In embedding coordinates, AdS$_{d+1}$ is the hypersurface
\be
-X_0^2 + X_1^2 + ... + X_{d}^2 - X_{d+1}^2 = - L^2 \; , \label{embed}
\ee
embedded in flat $(d+2)$-dimensional Minkowski space with two time directions. The AdS isometry group $O(2,d)$ is the Lorentz group of the embedding space and contains spatial rotations $M_{ij}$, two types of boosts, $M_{0i}$ and $M_{i (d+1)}$, as well as a rotation $M_{0 (d+1)}$ in the time-time plane. Consider irreducible rota-boosts of the form $M_{0 1} - \beta_1 M_{2 3} + ...$. There are two types of such boosts: those in which $X_0$ and $X_{d+1}$ are paired with $X_i$'s, and those in which $X_0$ and $X_{d+1}$ are paired with each other. In general a rota-boost of the first type with nonzero Casimir (\ref{Casimir}) can be written as
\be
M_{01} - \beta_1 M_{23} - \beta_2 M_{45} - ... \; .
\ee
Its norm is
\be
-X_1^2 + X_0^2 + \beta^2_1 \left(X_2^2 + X_3^2 \right) + ... + \beta^2_{d/2}\left(X_{d}^2 - X_{d+1}\right)^2 \; .
\ee
Using the embedding equation (\ref{embed}), this is not, for $d > 2$, time-like at the AdS boundary. Therefore, in higher dimensions, the above rota-boost cannot be considered as a candidate Hamiltonian. However, as we shall see, in three spacetime dimensions ($d  = 2$) a range of the parameter $\beta$ does give a timelike isometry, yielding the vacuum for rotating Rindler space.

\subsection{Rotating Rindler Space}

Consider the isometry generated by $M_{01} - \beta M_{23}$. This generator belongs to the loxodromic conjugacy class of rota-boosts. Technically, because we are dealing with the $AdS_3$ isometry group $SO(2,2)$, it is a combination not of a rotation and a boost but of two boosts in the embedding space:
\be
\frac{\partial}{\partial t} = \left(X_1\partial_0 + X_0\partial_1 \right) - \beta  \left(X_3\partial_2 + X_2\partial_3 \right) \; .
\ee
From the flat metric of the embedding space this has norm
\be
- (X_1^2 - X_0^2) (1- \beta^2) + \beta^2 \; ,
\ee
using the embedding equation. Restricted to the right Rindler wedge $X_1, X_1^2 - X_0^2 > 0$ we see that our candidate loxodromic generator is future-directed and timelike for $X_1^2 \gg X_0^2$ and has orbits that stay within the wedge. By construction, it is group-inequivalent to the usual (non-rotating) Rindler Hamiltonian $X_1 \partial_0 + X_0 \partial_1$, the invariant (\ref{Casimir}) for its conjugacy class being
\be
\epsilon_{0123}\omega^{01}\omega^{23} = 2\b \; .
\ee
The wedge admits a Cauchy surface on which one can define quantum states.

In 2+1 dimensions, rotating Rindler-AdS space can be coordinatized by
\bea
X_0 &=& \xi \sinh \left ( \frac{t}{L} - \b \frac{\chi}{L} \right ) \nonumber \\
X_1 &=& \xi \cosh \left ( \frac{t}{L} - \b \frac{\chi}{L} \right ) \nonumber \\
X_2 &=& \sqrt{L^2 + \xi^2} \sinh \left (\frac{\chi}{L} - \b \frac{t}{L} \right ) \nonumber \\
X_3 &=& \sqrt{L^2 + \xi^2} \cosh \left (\frac{\chi}{L} - \b \frac{t}{L} \right ) \; ,
\eea
where $\b$ is the rotation parameter. Here $-\infty < t, \chi < +\infty$ and $\xi > 0$.
The rotating Rindler metric is given by 
\be
ds^2 = -\left( (\xi/L)^2 (1 - \b^2) - \b^2 \right) dt^2  - 2\b dt \, d\chi + {d\xi^2 \over 1 + (\xi/L)^2} + \left(1 + (\xi/L)^2(1 - \b^2) \right) d\chi^2  \; . \label{Rotating}
\ee
The event horizon is at $\xi =0$; the determinant of the metric vanishes there. Notice also that at $\xi = \frac{\beta L}{\sqrt{1- \beta^2}}$, the t-t component of the metric vanishes. This indicates the presence of an ergosphere. Presumably this means that there are super-radiance effects in this space.

For $\beta = 0$, we recover the metric for non-rotating Rindler-AdS space:
\be
ds^2 = -(\xi/L)^2 dt^2  + {d\xi^2 \over 1 + (\xi/L)^2} + \left(1 + (\xi/L)^2 \right ) d\chi^2 \; . \label{NR-AdS}
\ee
Note that in the limit that the AdS radius, $L$, goes to infinity, so that $\xi/L \ll 1 $, the non-rotating metric gives ordinary (flat) Rindler space, where now $\frac{1}{L}$ is re-interpreted as the acceleration parameter of Rindler space instead of as the AdS scale. As a check we note that this limit is singular for the $\beta \neq 0$ metric, confirming that there is no rotating Rindler metric in flat space.

Both rotating and non-rotating Rindler-AdS space are of course a portion of anti-de Sitter space just as ordinary Rindler space is a piece of Minkowski space. In fact, even globally the portion of the spacetime covered by the coordinates above is identical to that covered by non-rotating Rindler 
coordinates. The diffeomorphism
\be
t \to t - \beta \chi	\qquad \chi \to \chi - \beta t
\ee
maps one spacetime to the other. In that sense, rotating Rindler space is classically the same spacetime as non-rotating Rindler space. However, the Hamiltonians for non-rotating and rotating Rindler space are isometrically-inequivalent and, as we shall see shortly, the corresponding vacuum states of scalar field theory are particle-inequivalent.

That rotating and non-rotating Rindler space describe the same part of spacetime may seem puzzling at first because one of them has an ergosphere and the other does not. This is because rotating Rindler-AdS space possesses an observer-dependent ergosphere, in addition to an observer-dependent event horizon. The existence of an observer-dependent ergosphere can be understood as follows. Recall the origin of the ergosphere for the Kerr black hole. In the two-dimensional subspace spanned by its time-translation and azimuthal Killing symmetries, the Kerr metric at large values of $r$ along the equator ($\theta = \pi/2$) approaches $-dt^2 + r^2 d\phi^2$, because the Kerr spacetime is asymptotically flat. Therefore, for the Kerr black hole there is a unique choice of Killing vector that is timelike at infinity, namely $(d/dt)^a$; any other linear combination of $(d/dt)^a$ and $(d/d \phi)^a$ is spacelike at infinity. The Killing vector corresponding to time translations is therefore fixed, and hence so is the place where it becomes null i.e. the ergosphere. The geometry ensures that the location of the ergosphere is unambiguous. Contrast this with the situation in AdS. The metric for the two-space spanned by the time-translation and azimuthal Killing vectors in Rindler-AdS approaches $(\xi/L)^2 (-dt^2 + d \chi^2)$. The boundary metric is simply a re-scaled two-dimensional Minkowski metric. Any observer moving along a timelike linear combination of $(d/dt)^a$ and $(d/d \chi)^a$ can choose his or her worldline as the time-translation direction. Each such linear combination of Killing vectors becomes null in a different place. Consequently, the existence and location of the ergosphere are both observer-dependent.

For each of the different possible time choices labeled by $\beta$, there is a corresponding stationary vacuum state annihilated by the Hamiltonian that generates that time evolution. We shall call this one-parameter family of vacuum states ``$\beta$-vacua." Like the $\alpha$-vacua of de Sitter space \cite{Allen}, these vacuum states are particle-inequivalent. The particle-inequivalence of the $\beta$-vacua to the usual Rindler vacuum (and to each other) can be verified explicitly by a Bogolubov transformation. Consider a positive-frequency ($\omega > 0$) mode of the Klein-Gordon equation:
\be
u_{k, \omega} (t, \chi, \xi) = e^{-i\omega t + ik\chi} f_{\omega, k}(\xi) \; . \label{u-mode}
\ee
Demanding normalizability with respect to the Klein-Gordon inner product, one can show \cite{Esko} that the value of $\omega$ does not constrain $k$. Now, under the transformation $t \rightarrow t - \beta \chi$ and $\chi \rightarrow \chi - \beta t$, the mode transforms as 
\be
u_{k, \omega} \rightarrow e^{-i(\omega + \beta k) t}e^{i(k + \beta \omega)\chi}f_{\omega, k} (\xi) \; .
\ee
We see that for $k<-\frac{\omega}{\beta}$, the mode has negative frequency. Hence there is a mixing between the negative and positive frequency modes under transformation from rotating to a non-rotating Rindler-AdS space. This fact can be formally demonstrated in terms of the Bogolubov coefficients. Consider a positive frequency mode with respect to one of the $\beta$ rota-boosts:
\be
v_{l,\nu} = e^{-i\nu t' + il\chi'} g_{l,\nu} (\xi) \; . \label{v-mode}
\ee
Since $t = t' - \beta \chi'$ and $\chi = \chi' - \beta t'$, (\ref{v-mode}) can be re-expressed in terms of the modes of the nonrotating vacuum:
\be 
u_{k',\omega'} = e^{-i\omega' t + ik'\chi}  f_{k',\omega'} (\xi) \; ,
\ee  
where $\omega' = \frac{\nu - \beta l}{1 - \beta^2}$ and $k' = \frac{l - \beta \nu}{1 - \beta^2}$. If $\omega'<0$, then the beta Bogolubov coefficient is nonzero between (\ref{u-mode}) and (\ref{v-mode}) and can be easily calculated as \cite{Davies}
\be
\beta(k,\omega;l,\nu) =  i\Theta(-\nu + \beta l) \delta \left(\omega + \frac{\nu - \beta l}{1 - \beta^2}\right) \delta \left(k + \frac{l - \beta \nu}{1 - \beta^2}\right) \; , \label{Bogolubov} 
\ee  
while the Bogolubov alpha coefficient vanishes. 

The expression for the beta coefficient implies that the nonrotating Rindler observer perceives any $\beta$-vacuum as filled with an infinite sea of particles for each positive frequency $\omega$. Of course the global AdS vacuum appears thermal with a different temperature for each $\beta$-vacuum observer. Indeed, already from the metric it is clear that the temperature depends on the rotation parameter, $\beta$:
\be
T = \frac{1 - \beta^2}{2 \pi L} \; .
\ee
Interestingly, the limit $\beta \to 1$ appears to correspond to an extremal vacuum state in Rindler-AdS space, with vanishing temperature. 

\subsubsection{Rotating Rindler space and the BTZ black hole}
The existence of an ergosphere in AdS space recalls the rotating BTZ black hole. Indeed, Rindler-AdS space is related to the BTZ black hole \cite{BTZ} via 
\be
\chi \sim \chi + 2\pi a \; .
\ee
A change of coordinates
\be
\xi = \sqrt{\f{r ^2 - 1}{1 - \b ^2}}
\ee
puts the metric in the familiar BTZ form:
\be
ds^2 = -\frac{(r ^2 - 1)(r ^2 - \b ^2)}{r^2}dt^2 + \f{r^2}{(r ^2 - 1)(r ^2 - \b ^2)}dr^2 + r^2 \left (d\chi - \f{\b}{r^2}dt \right )^{\! 2} \; .
\ee
Rindler-AdS (\ref{NR-AdS}) is the universal cover for the BTZ black hole \cite{Emparan,Myers,Lowe,Lowe2,Vanzo}. The black hole solution is obtained by making an identification in a direction perpendicuar to $\partial_t$ at the boundary. However, there is an important difference between Rindler-AdS space and the BTZ black hole. The identification breaks the symmetry group down from $SL(2,R) \times SL(2,R)$ to $SL(2,R) \times U(1)$. Consequently, the freedom of picking out the time direction is lost; neither the event horizon nor the ergosphere of the BTZ black hole is observer-dependent. Put another way, the identification $\chi \sim \chi + 2 \pi a$ gives the two-dimensional boundary Minkowski space a cylinder topology. But special relativity on a cylinder has a preferred frame, singled out by the identification \cite{BarrowLevin,Greene}. Hence there is a preferred direction of time.

\subsection{Rotating Global Vacua}

Another type of loxodromic generator in AdS is
\be
\frac{\partial}{\partial t} = \left(X_0\partial_3 - X_3\partial_0 \right) - \beta  \left(X_1\partial_2 - X_2\partial_1 \right) \label{loxglobal} \; .
\ee
This is a combination of a temporal and a spatial rotation. For comparison, the generator of global time, $\tau$, is just the temporal rotation $\left(X_0\partial_3 - X_3\partial_0 \right)$.

The embedding coordinates can be parameterized by
\bea
X_0 &=& \sqrt{\f{r^2 + 1}{1 - \b^2}} \cos ( t - \b \theta) \nonumber \\
X_3 &=& \sqrt{\f{r^2 + 1}{1 - \b^2}} \sin ( t - \b \theta) \nonumber\\
X_1 &=& \sqrt{\f{r^2 + \b^2}{1 - \b^2}} \cos(\theta - \b t) \nonumber \\
X_2 &=& \sqrt{\f{r^2 + \b^2}{1 - \b^2}} \sin(\theta - \b t) \; .
\eea
Then, with the AdS scale set to unity, the line element reads
\bea
ds^2 = -\left(r^2 + 1 + \beta^2 \right) dt^2 + \frac{r^2 dr^2}{(1 + r^2)(\b^2 + r^2)} + r^2 d\theta^2 + 2\beta dtd\theta \; . \label{betaglobal}
\eea
Here we have $0 \leq \beta<1$, $0 \leq r < \infty$, and $\theta \sim \theta + 2\pi$. Clearly when $\beta = 0$ this reduces to the AdS metric in global coordinates, as it should.

For $\beta \neq 0$, there is however a subtlety with this solution. The generator of rotations $\frac{\partial}{\partial \theta}$ in embedding coordinates is
\be
\frac{\partial}{\partial \theta} = -\beta \left(X_0\partial_3 - X_3\partial_0 \right) +  \left(X_1\partial_2 - X_2\partial_1 \right) \; . \label{loxglobaltheta}
\ee
This has norm $-\beta^2 + (X_1^2 + X_2^2) (1-\beta^2)$, which, however, becomes timelike for
\be
X_1^2 + X_2^2 = \frac{r^2 + \beta^2}{1-\beta^2} \leq \frac{\beta^2}{1-\beta^2} \; .
\ee
So that region cannot be covered by this coordinate system. To see where that region is, we note that the relation between the radius in global coordinates, $\rho$, and $r$ is
\be
\rho^2 = \frac{r^2 + \beta^2}{1-\beta^2} \; .
\ee
For $\beta \neq 0$, $r =0$ no longer corresponds to the center $\rho = 0$ of the AdS cylinder but is instead a surface of non-zero $\rho$. That is, we have effectively removed a concentric cylinder from within the AdS cylinder for the purposes of this coordinate system. 

At AdS infinity, however, nothing has changed and so we can calculate the conserved charges of this space. The mass and angular momentum of rotating global AdS space can be evaluated using the prescription of \cite{VJ-Kerr-dS}. The result is
\bea
M &=&- \frac{1}{8 \pi G} \int_0 ^{2\pi} \f{r^4}{2}\left(\frac{1+\beta^2}{r^4} \right)d\theta = -\frac{1+\beta^2}{8G} \\
|J| &=& \frac{1}{8\pi G}\int_0 ^{2\pi} \beta d\theta=\frac{\beta}{4G} \; .
\eea

Of course we cannot actually remove the inner region because then the spacetime would be geodesically incomplete. However, we can still do quantum field theory in the region outside the inner cylinder using (\ref{loxglobal}) for time evolution. (The presence of the inner cylinder means that the spacetime does not admit a Cauchy surface; however, global AdS space already has this problem so, presumably, this is not much worse.) A similar problem afflicts Kerr-de Sitter space in three dimensions \cite{VJ-Kerr-dS}. We can nevertheless circumvent the problem by defining the angular generator to be $\frac{\partial}{\partial \theta} = \left(X_1\partial_2 - X_2\partial_1 \right)$. Unlike (\ref{Rotating}), the Killing vectors $\pa_{\theta}$ and $\pa_t$ would then not be orthogonal to each other at the conformal boundary. The line element can be written as
\be
- \left(1 + r^2 (1 - \b^2) \right) dt^2 + \frac{dr^2}{1 + r^2} + r^2 d\theta^2 - 2\b r^2 dt \, d\theta  \; . \label{globalbeta}
\ee 
   
To show that the $\beta$-vacua corresponding to the time choice (\ref{loxglobal}) are distinct from the global AdS vacuum, we need to again show that positive and negative frequency modes mix. Normalizability conditions for fields in global AdS space were investigated in \cite{VJ-Global-AdS}. Using the Ansatz $\Phi (r,t,\theta) = e^{-i\omega t}e^{im\theta}f(r)$ (where $m \in Z$) in global coordinates, Klein-Gordon normalizability implies that 
\be
\omega = \pm \left|2h_{+} + m + 2n \right|\; , ~~n = 0,1,2, ...  \; , \label{quantcondition}
\ee
where $h_{+} = 1 + \sqrt{1 + M^2}$ with $M$ the mass of the scalar field. Under the transformation $\theta \rightarrow \theta - \beta t$, which takes the global coordinates metric into (\ref{globalbeta}), the mode solutions become
\be 
\Phi \rightarrow e^{-i(\omega + \beta m) t}e^{i m \theta}f(r) \; .
\ee
Given any value of $n$, we see from (\ref{quantcondition}) that we can always find sufficiently large values of negative $m$ such that $\omega < |\beta m|$. A positive frequency mode can therefore become a negative frequency mode, and hence the rotating global AdS $\beta$-vacua are different from the global AdS vacuum. This can also be confirmed by calculating the Bogolubov coefficients directly as we did for rotating Rindler space. In higher dimensions, the quantization condition becomes \cite{VJ-Global-AdS}
\be
\omega = \pm \left|2h_{+} + l + 2n \right|;~~l,n = 0,1,2, ... \; .
\ee
By the semi-positiveness of $l$, we always have $\omega> l$ and hence the positive frequency and negative frequency modes cannot mix under a transformation to rotating coordinates. $\beta$-vacua therefore do not exist in higher-dimensional global AdS space.

The transformation between rotating and non-rotating coordinates can also be studied from the boundary theory. The global isometries of $AdS_3$ become the Virasoro generators at the boundary:
\be
\begin{array}{rclrcl}
l_0 & =  & i\pa_w & \qquad \bar{l}_0 & = & i\pa_{\bar{w}} \\
l_{+1} & = & ie^{+iw}\pa_w & \qquad \bar{l}_{+1} & = & ie^{+i\bar{w}}\pa_{\bar{w}} \\
l_{-1} & = & ie^{-iw}\pa_w & \qquad \bar{l}_{-1} & = & ie^{-i\bar{w}}\pa_{\bar{w}}
\end{array}
\ee
where $w=t+\theta$ and $\bar{w}=t-\theta$. The transformation to rotating global AdS implies $w \rightarrow (1 - \beta)w$ and $\bar{w} \rightarrow (1 + \beta)\bar{w}$. With this, the generators become
\be
\begin{array}{rclrcl}
l'_0 & = & \frac{i}{1-\b} \pa_w & \qquad \bar{l}'_0 & = & \f{i}{1+\b}\pa_{\bar{w}} \\
l'_{+1} & = & \f{i}{1-\b} e^{+i(1-\b)w}\pa_w & \qquad \bar{l}'_{+1} & = & \f{i}{1+\b} e^{+i(1+\b)\bar{w}}\pa_{\bar{w}} \\
l'_{-1} & = & \f{i}{1-\b} e^{-i(1-\b)w}\pa_w & \qquad \bar{l}'_{-1} & = & \f{i}{1+\b} e^{-i(1+\b)\bar{w}}\pa_{\bar{w}}
\end{array}
\ee
These also satisfy the Virasoro algebra. However, the transformation from non-rotating to rotating global AdS is not a conformal transformation and therefore does not preserve the conformal vacuum at the boundary.

\section{Discussion}
By considering the conjugacy classes of AdS isometry group, which is a higher-dimensional Lorentz group, we have been led to find a class of inequivalent vacua in three-dimensional anti-de Sitter space. The same line of reasoning can also be applied to de Sitter space, but because de Sitter space does not have a spatial boundary, the conditions for obtaining stationary vacua are somewhat unclear. It would be worthwhile to pursue this further. We have also shown that the $\beta$-vacua are obtained from the nonrotating Rindler vacuum via a nontrivial Bogolubov transformations of the mode functions. A related problem would be to calculate the response of an Unruh detector moving along orbits of $\partial_t$ in rotating Rindler-AdS space \cite{DeserLevin1,DeserLevin2}. Moreover, since the rotating Rindler-AdS vacuum becomes extremal in the limit $\beta \rightarrow 1$, it might be that such a vacuum is supersymmetric. Another potentially interesting direction for future would might be to consider rotating vacuum states in the context of AdS/CFT applied to Rindler-AdS space \cite{Hawking-Robinson,berman,RAdS}. Finally, the existence of an observer-dependent ergosphere raises an amusing question. Spacetimes with ergospheres generally allow for super-radiance effects. For Kerr black holes, the effect of super-radiant scattering is to ``unwind" the black hole, by depleting its angular momentum. Rotating Rindler space however has infinite angular momentum. Is it possible then to have eternal super-radiance? That would not necessarily be problematic; indeed, Rindler space can also Hawking-radiate eternally. It would be interesting to develop these ideas.

\bigskip
\noindent
{\bf Acknowledgments}

\noindent
We would like to thank Vijay Balasubramanian, Steven Carlip, and Paul Davies for helpful discussions. M.~P. is supported in part by DOE grant DE-FG02-09ER41624.


\begin{thebibliography}{10}
\providecommand{\url}[1]{\texttt{#1}}
\providecommand{\urlprefix}{URL }
\expandafter\ifx\csname urlstyle\endcsname\relax
  \providecommand{\doi}[1]{doi:\discretionary{}{}{}#1}\else
  \providecommand{\doi}{doi:\discretionary{}{}{}\begingroup
  \urlstyle{rm}\Url}\fi
\providecommand{\eprint}[2][]{\url{#2}}

\bibitem{Davies}
 N.~D.~Birrell, and P.~C.~W.~Davies, 
\textit{Quantum fields in curved space},
Cambridge (1982).

\bibitem{Pfautsch}
 J.~R.~Letaw and J.~D.~Pfautsch, 
``The Quantized Scalar Field in the Stationary Coordinate Systems of Flat Space-Time,"
Phys. Rev. D {\bf 24} (1981) 1491.


\bibitem{daviesrotating}
P.~C.~W.~Davies, T.~Dray, and C.~A.~Manogue, ``Detecting the rotating quantum vacuum," Phys. Rev. D {\bf 53} (1996) 4382.


\bibitem{Allen}
B.~Allen, ``Vacuum States in de Sitter Space," Phys. Rev. D {\bf 32} (1985) 3136.

\bibitem{Esko}
 E.~Keski-Vakkuri,
``Bulk and boundary dynamics in BTZ black holes,"
Phys. Rev. D {\bf 59} (1999) 104001; {\tt hep-th/9808037}.


\bibitem{BTZ}
 M.~Banados, C.~Teitelboim, and  J.~Zanelli,
``The black hole in three-dimensional space-time,"
Phys. Rev. Lett. {\bf 69} (1992) 1849-1851; {\tt hep-th/9204099}.


\bibitem{Emparan}
 R.~Emparan,
``AdS/CFT duals of topological black holes and the
entropy of zero energy states,"
JHEP {\bf 9906} (1999) 036; {\tt hep-th/9906040}.


\bibitem{Myers}
 R.~C.~Myers, and  A.~Sinha,
``Seeing a c-theorem with holography,"
Phys. Rev. D {\bf 82} (2010) 046006; {\tt arXiv:1006.1263}.


\bibitem{Lowe}
 A.~Hamilton,  D.~N.~Kabat,  G.~Lifschytz, and D.~A.~Lowe,
``Holographic representation of local bulk operators,"
Phys. Rev. D {\bf 74} (2006) 066009; {\tt hep-th/0606141}.


\bibitem{Lowe2}
A.~Hamilton,  D.~N.~Kabat,  G.~Lifschytz, and D.~A.~Lowe,
``Local bulk operators in AdS/CFT correspondence: A boundary view of horizons and locality,"
Phys. Rev. D {\bf 73} (2006) 086003; {\tt hep-th/0506118}.


\bibitem{Vanzo}
 L.~Vanzo, ``Black holes with unusual topology,"
Phys. Rev. D {\bf 56} (1997) 6475; {\tt gr-qc/9705004}.

  
\bibitem{BarrowLevin}
J.~D.~Barrow and J.~J.~Levin, ``The twin paradox in compact spaces," Phys. Rev. A {\bf 63} (2001) 044104; {\tt gr-qc/0101014}.


\bibitem{Greene}
 B.~Greene,  J.~Levin, and  M.~Parikh,
``Brane-World Motion in Compact Dimensions,"
Class. Quant. Grav. {\bf 28} (2011) 155013; {\tt arXiv:1103.2174}.


\bibitem{VJ-Kerr-dS}
 V.~Balasubramanian,  J.~de Boer, and  D.~Minic,
``Mass, entropy and holography in asymptotically de Sitter spaces,"
Phys. Rev. D {\bf 65} (2002) 123508; {\tt hep-th/9805171}.


\bibitem{VJ-Global-AdS}
V.~Balasubramanian, P.~Kraus, and A.~E.~Lawrence,
``Bulk vs. boundary dynamics in anti-de Sitter spacetime,"
Phys. Rev. D {\bf 59} (1999) 046003; {\tt hep-th/0110108}.


\bibitem{DeserLevin1}
S.~Deser and O.~Levin,
``Accelerated detectors and temperature in (anti)-de Sitter spaces,"
Class. Quant. Grav. {\bf 14} (1997) 064004; {\tt gr-qc/9706018}.


\bibitem{DeserLevin2}
S.~Deser and O.~Levin,
``Mapping Hawking into Unruh thermal properties,"
Phys. Rev. D {\bf 59} (1999) 064004; {\tt hep-th/9809159}.


\bibitem{Hawking-Robinson}
S.~W.~Hawking, C.~J.~Hunter, and M.~Taylor,
``Rotation and the AdS/CFT correspondence,"
Phys. Rev. D {\bf 59} (1999) 064005; {\tt hep-th/9811056}.


\bibitem{berman}
D.~S.~Berman and M.~K.~Parikh, ``Holography and rotating AdS black holes," Phys. Lett. B {\bf 463} (1999) 168; {\tt hep-th/9907003}.


\bibitem{RAdS}
M.~Parikh, P.~Samantray, and E.~Verlinde, to appear.


\end{thebibliography}
\end{document}